\documentclass[]{jpp}

\usepackage{graphicx}
\usepackage[utf8]{inputenc}

\usepackage{amsmath}
\usepackage{amsfonts}
\usepackage{color,xcolor}
\usepackage{mathrsfs,amsmath}
\usepackage{amssymb,color}
\usepackage{mathtools}
\usepackage{graphicx}
\usepackage{aas_macros}
\usepackage{array}
\usepackage{url}
\usepackage{longtable}

\usepackage{hyperref}

\usepackage{multirow}
\usepackage{amssymb}

\usepackage[normalem]{ulem} 

\usepackage[mathscr]{euscript}

\newcommand{\mathbfit}[1]{\textbf{\textit{#1}}}
\renewcommand{\vec}[1]{\mathbfit{#1}}
\newcommand{\bs}[1]{\boldsymbol{#1}}
\newcommand{\A}{\mathrm{A}}

\newcommand{\alf}{Alfv$\acute{\text{e}}$n} 

\DeclareSymbolFont{matha}{OML}{txmi}{m}{it}
\DeclareMathSymbol{\varv}{\mathord}{matha}{29}

\SetSymbolFont{symbols}{bold}{OMS}{cmsy}{b}{n}
\DeclareSymbolFont{bmisymbols}{OML}{cmm}{b}{it}
\DeclareMathSymbol{\bvarv}{0}{bmisymbols}{"1D}

\usepackage{array}
\newcolumntype{M}[1]{>{\centering\arraybackslash}m{#1}}
\newcolumntype{N}{@{}m{0pt}@{}}

\usepackage{amsmath}
\usepackage{xcolor}
\usepackage{natbib}
\usepackage{xpatch}

\usepackage{hyperref}
\hypersetup{
	colorlinks=true,
	breaklinks=true,
	allcolors=blue,
	frenchlinks=true
}

\makeatletter
\xpatchcmd\NAT@citex
{%
	\@citea\NAT@hyper@{%
		\NAT@nmfmt{\NAT@nm}%
		\hyper@natlinkbreak{\NAT@aysep\NAT@spacechar}{\@citeb\@extra@b@citeb}%
		\NAT@date
	}%
}
{%
	\@citea
	\NAT@nmfmt{\NAT@nm}%
	\NAT@aysep\NAT@spacechar
	\NAT@hyper@{\NAT@date}%
}
{}{}
\xpatchcmd\NAT@citex
{%
       \@citea\NAT@hyper@{%
               \NAT@nmfmt{\NAT@nm}%
               \hyper@natlinkbreak{\NAT@spacechar\NAT@@open\if*#1*\else#1\NAT@spacechar\fi}%
               {\@citeb\@extra@b@citeb}%
               \NAT@date
       }%
}
{
       \@citea
       \NAT@nmfmt{\NAT@nm}%
       \NAT@spacechar\NAT@@open\if*#1*\else#1\NAT@spacechar\fi
       \NAT@hyper@{\NAT@date}%
}
{}{}
\makeatother

\newcommand{\rmn}{\mathrm}

\title{
Deciphering the physical basis of the intermediate-scale instability
}

\author{
Mohamad Shalaby
\corresp{\email{mshalaby@live.ca}},
Timon Thomas,
Christoph Pfrommer,\\
Rouven Lemmerz,
and
Virginia Bresci
}

\shorttitle{Physics of the intermediate-scale instability}
\shortauthor{Shalaby et al.}

\affiliation{Leibniz-Institut f{\"u}r Astrophysik Potsdam (AIP), An der Sternwarte 16, D-14482 Potsdam, Germany}

\begin{document}

\maketitle

\begin{abstract}
We study the underlying physics of cosmic-ray (CR) driven instabilities that play a crucial role for CR transport across a wide range of scales, from interstellar to galaxy cluster environments. By examining the linear dispersion relation of CR-driven instabilities in a magnetised electron-ion background plasma, we establish that both, the intermediate and gyroscale instabilities have a resonant origin and show that these resonances can be understood via a simple graphical interpretation. These instabilities destabilise wave modes parallel to the large-scale background magnetic field at significantly distinct scales and with very different phase speeds. Furthermore, we show that approximating the electron-ion background plasma with either magnetohydrodynamics (MHD) or Hall-MHD fails to capture the fastest growing instability in the linear regime, namely the intermediate-scale instability. This finding highlights the importance of accurately characterising the background plasma for resolving the most unstable wave modes. Finally, we discuss the implications of the different phase speeds of unstable modes on particle-wave scattering. Further work is needed to investigate the relative importance of these two instabilities in the non-linear, saturated regime and to develop a physical understanding of the effective CR transport coefficients in large-scale CR hydrodynamics theories.
\end{abstract}

\section{Introduction}
\label{sec:intro}

The majority of astrophysical plasmas are likely to be permeated with CRs; these include protoplanetary disks, the interstellar, circumgalactic and intracluster media.
In the Milky Way, the CR energy density (dominated by a CR population at around GeV energies) is in equipartition with the average thermal and magnetic energy densities \citep{Boulares1990}.
Hence, these CRs constitute an essential nonthermal component that provides dynamical feedback to the interstellar medium \citep{2016Girichidis,2018Girichidis,2016Simpson,2023Simpson,2018Farber} and can launch galaxy-scale outflows as demonstrated in one-dimensional models \citep{1975Ipavich,1991Breitschwerdt,2016Recchia,2022Quataert} as well as in three-dimensional simulations of galaxies forming both in isolation \citep{2012Uhlig,2014Salem,2016PakmorIII,2017Ruszkowski,2023Thomas} and in cosmological environments \citep{2014SalemII,2020Hopkins,2020Buck}.
The propagation of CRs with energies below 100 GeV is believed to be predominantly governed by self-generated magnetic perturbations \citep{2012Blasi, 2018Evoli}.
These perturbations efficiently scatter the CRs, resulting in a significant decrease in their mean transport speed. The interplay of CR-driven growth of plasma waves and collisionless wave damping processes determines the effective transport speed of CRs and hence, the coupling strengths to the ambient plasma: strong scattering causes CR isotropisation in the {\alf} wave frame and forces CRs to stream at mean speeds close to the {\alf} speed while faster CR diffusion prevails in the case of weak scattering if the waves are strongly damped \citep{Zweibel2017}.

Most importantly, the strength of CR feedback critically depends on these microscopic transport properties, including mass and energy loading factors of galactic winds, the wind speed and the emerging CR pressure support in the circumgalactic medium \citep{2020Buck,2020Ji}. Likewise, the evolution of gas and chemical compositions in diffusive regions and dense interstellar clouds are affected by CR ionisation to the extent that CRs are the fundamental source of residual ionisation inside shielded molecular clouds \citep{2018Phan}. Because of the sensitive dependence of the emerging galaxies and the phase structure of the interstellar medium on CR feedback strength, the investigation of CR-driven instabilities becomes of paramount importance in order to attain predictive capabilities in simulation campaigns. Thus, understanding and accurately modelling these instabilities is crucial for effectively regulating the transport of CRs in interstellar, circumgalactic, and intracluster plasmas.

Our focus here is on instabilities of parallel wave modes along the background magnetic field.
This is for two reasons, which follow for astrophysical plasmas, where the CR density is much lower compared to the background plasma.
First, when considering the full spectra of CR driven waves, the growth rates due to resonance with hydrodynamical waves are highest for parallel wave mode, as demonstrated in equation~(4) of \citet{kulsrud+1969}.
Second, even if oblique waves
which propagate at an angle to the magnetic field, were to grow, these waves would be strongly Landau damped by thermal background ions, especially in high plasma beta conditions (where the thermal energy is  greater than the magnetic energy). 
\citet{Foote+1979} estimated the damping rate and showed that it is significantly faster than the typical growth rates of obliquely propagating wave modes (see equation (68) of \citealt{Zweibel2017} for a concise expression of the damping rate).

In CR-driven instabilities, CR ions represent the primary source of free energy. This energy is channelled through 
these instabilities into unstable electromagnetic wave modes of the background plasma.
This is in contrast to beam-plasma instabilities \citep{breizman+1972,bret-2010-pop,chang:2014,resolution-paper}, where the source of free energy is an electron-positron beam that drives wave modes on the electron skin depth scale or shorter unstable.
By contrast, CR-driven instabilities excite unstable electromagnetic wave modes on scales much larger than the electron skin depth. These instabilities can be classified generally as follows.
First, there are non-resonant instabilities, such as the Bell instability \citep{Bell2004}, which occur when the CR current is very high. These instabilities are highly relevant for studying the escape of CR ions after acceleration at supernova remnant shocks.
Second, there are resonant instabilities, which occur when the CR current is low but the CR mean drift is faster compared to the local ion {\alf} speed.
Within galactic and stellar environments, CR transport (with energies below 100 GeV) is believed to be primarily regulated by two dominant resonant instabilities: the gyro-resonant instability \citep{kulsrud+1969,2018Lebiga,HS+2019,Bai2019, bai2021, bambic2021, plotnikov2021} and the recently discussed intermediate-scale instability \citep{sharp2,Lemmerz+2023}.
Provided CRs propagate with a finite pitch angle relative to the large-scale magnetic field, they induce an instability in electromagnetic waves (propagating along the background magnetic field) on scales intermediate between the gyroradii of ions and electrons. This instability occurs as long as CRs drift at velocities less than half of the  {\alf} speed of electrons. The emerging unstable modes are identified as background ion-cyclotron modes in the reference frame co-moving with the CRs.  Interestingly, this newly found instability typically exhibits significantly faster growth, exceeding the growth rate of the commonly discussed resonant instability at the ion gyroscale by more than an order of magnitude \citep{sharp2}.

We focus, moreover, on instabilities in the cold limit for background plasma species. While it remains to be shown analytically and numerically that this assumption does not impact the nature of the instabilities we study in this paper, the unstable wave modes are typically present on scales much larger than the electron skin depth (which is typically larger than both ion and electron Debye lengths) and of electromagnetic nature. Thus, background temperatures are typically argued to have no impact on such  long-wavelength unstable wave modes \citep{Zweibel-2003,Bell2004,sharp2}.
This is in contrast to beam-plasma instabilities that can be greatly impacted by thermal effects \citep{bret-2010-pre,linear-paper,shalabythesis2017} or structures \citep{krafft-2013,sim_inho_18,th_inho_20} in the background plasma on scales close to the electron skin-depth.
The intermediate-scale instability can also play a significant role in electron acceleration within non-relativistic shocks \citep{Shalaby+2022ApJ}.
This study focuses on unravelling the physical origins of these resonant instabilities, particularly in the context of CRs with a gyrotropic momentum distribution. Through a transparent visualisation of the fundamental mechanisms, we gain a deeper understanding of the behaviour and characteristics of these instabilities and their influence on CR transport.

The structure of the paper is as follows. First, in Section~\ref{sec:dispersions}, we analyse the normal modes supported by a magnetised electron-ion plasma system and investigate how these wave modes are affected when an additional population of electron-ion CRs, characterised by a relative drift speed, is included.
This inclusion leads to the emergence of additional Doppler-shifted CR wave modes, and the interaction between the background and CR ion wave modes gives rise to resonant instabilities, with the maximum growth rate occurring at the wavelengths where these modes resonantly interact.
Next, in Section~\ref{sec:Bgs}, we demonstrate that approximating the background plasma using either MHD or Hall-MHD erroneously overlooks the fastest growing modes in the linear regime. This highlights the importance of considering the full dynamics of the system down to the electron scale.
Furthermore, in Section~\ref{sec:scattering}, we explore the impact of the driven wave modes on particle-wave scattering and how the presence of these wave modes influences the particle trajectories.
Finally, in Section~\ref{sec:summary} we summarise our findings and present an outlook for potential implications in Section~\ref{sec:outlook}.
Throughout this work, we use the SI system of units.

\section{Electromagnetic linear dispersion relation}
\label{sec:dispersions}

In a magnetised plasma with various species $s$, the linear dispersion relation for electromagnetic wave modes of (complex) frequency $\omega$ and wave mode $k$ that propagate parallel to a constant background magnetic field (here taken to be $\vec{B}_0 = B_0 \bs{\hat{x}}$) is given by \citep{Schlickeiser+2002}%

\begin{eqnarray}
D^{\pm} &=& \omega^2 - k ^2c^2  +  \sum_s \chi^{\pm}_s = 0,
~~
{\rm where,}~~
\\
\chi^{\pm}_s
&=&
\frac{ q^2_s }{m_s  }
\!\int\!\! d^3u
 \frac{f_{s,0}  (u_{\parallel} , u_{\perp}) }{\gamma}
\left[
\frac{\omega -k \varv_{\parallel}}{k \varv_{\parallel}-\omega \pm \Omega_s }
-
\frac{\varv_{\perp}^2 c^{-2}\left(k^2c^2-\omega ^2  \right)}
{2 \left(k \varv_{\parallel} - \omega  \pm 
\Omega_s
\right){}^2}
\right].
\end{eqnarray}
Here, $\chi_s^{\pm}$ is the linear response for species $s$, which is characterised by charge $q_s$, mass $m_s$, and the equilibrium gyrotropic phase-space distribution, $f_{s,0}$. 
The light speed in vacuum is denoted by $c$ and the spatial part of the 4-velocity is $\vec{u} =\gamma \bvarv$, where, $u_{\parallel}$ and $u_{\perp}$  are the parallel and perpendicular velocity, respectively, which are defined with respect to the direction of the uniform background magnetic field $\vec{B}_0$. The magnitude of the velocity is defined such that $u^2 = u_{\parallel}^2 + u_{\perp}^2$ and $\gamma=\sqrt{1+u^2/c^2}$.
The non-relativistic cyclotron frequency of species $s$ is
$\Omega_{s,0}  = q_s B_0/ m_s   $,  and the relativistic one is $\Omega_s =  \Omega_{s,0} / \gamma $.

\begin{figure}
\centering
\includegraphics[width=8cm]{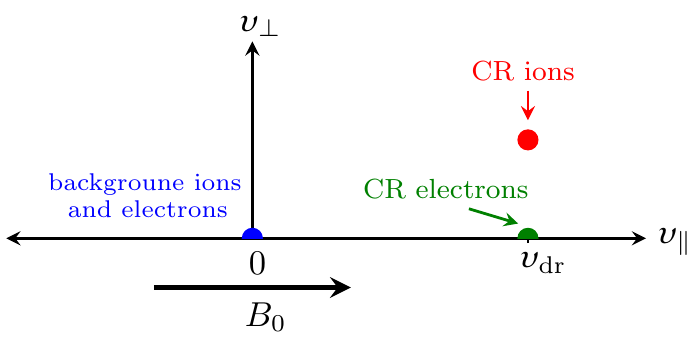}
\caption{\label{fig:fig04}%
Schematic representation of the velocity distributions adopted for our analytical calculation in this paper. We chose a cold distribution for background electrons and ions, and a gyrotropic distribution for CR electrons and ions. While a gyrotropic CR distribution is adopted here for analytical tractability, the nature of resonances that lead to the instabilities discussed in this work is independent of such a choice.
}
\end{figure}

For simplicity, we take  $f_{s,0} $ to be gyrotropic ring distribution with fixed parallel and perpendicular velocities for all particles of species $s$, i.e., $f_{s,0}(u_{\parallel},u_{\perp}) = n_s \delta(u_{\parallel} - \gamma \varv_{\rm dr}) \delta(u_{\perp} - \gamma \varv_{\perp})/ (2 \pi u_{\perp})$, where, $n_s$ is a uniform density of the plasma species $s$, and all of its particles are drifting along $\vec{B}_0$ with constant speed $ \varv_{{\rm dr},s}$ and have the same perpendicular velocity $\varv_{\perp,s}$. In this case, the linear response reduces to $\zeta_s^{\pm} $ given by \citep{HS+2019,Weidl2019,sharp2}
\begin{align}
\chi_s^{\pm}  \rightarrow
\zeta_s^{\pm} ( \varv_{{\rm dr},s},\varv_{\perp,s}
,n_s)
&=
\frac{ \omega^2_s }{ \gamma} 
\left[
\frac{\omega -k \varv_{{\rm dr},s} }{k \varv_{{\rm dr},s} -\omega \pm \Omega_s }
-
\frac{\varv_{\perp,s} c^{-2}\left(k^2c^2-\omega ^2  \right)}
{2 \left(k \varv_{{\rm dr},s}  - \omega  \pm \Omega_s\right){}^2}
\right].
\label{eq:DispGyro}
\end{align}
Here, $\omega_{s} = \sqrt{ n_{ s} q^2_{s}/(m_{ s} \epsilon_0) }$ is the plasma frequency of species $s$ and $\epsilon_0$ is the permittivity of free space. 
The solutions of $D^{\pm}=0$ correspond to  right/left  polarisation states, and the corresponding first-order perturbed electric and magnetic fields obey
$ E_{y,1}(k,\omega) \pm i E_{z,1}(k,\omega) =0$ and
$ B_{y,1}(k,\omega) \pm i B_{z,1}(k,\omega) =0$ respectively.
In the analysis below, the velocity distributions for both the background plasma and cosmic ray (CR) species are graphically illustrated in Figure~\ref{fig:fig04}; a cold distribution for background electrons and ions and a gyrotropic distribution for CRs.

\subsection{Normal modes of electron-ion plasma}

The normal modes of an electron-ion plasma are well known and extensively studied in the literature \citep[see, e.g.,][]{boyd}.
In the cold-limit (also called the cold-hydrodynamic (HD) limit), one can obtain the plasma rest-frame wave modes by solving the following dispersion relation:
\begin{eqnarray}
\label{eq:Bgdisp}
\omega^2- k^2 c^2 + \zeta_{\rm e}^{-} (0,0,n_0) + \zeta_{\rm i}^{-} (0,0,n_0) = 0, 
\end{eqnarray}
where $\zeta_{\rm i(e)}$ denotes the contribution of the cold stationary ion (electron) species with a fixed uniform number density $n_0$.
Therefore, for $k>0$, the real part of the solutions, $\omega_{\rm r}=\Real(\omega)>0 ~ (<0)$ indicate that the sense of rotation of the  magnetic eigenmodes is the same as that of electrons (ions) around the constant background magnetic field $\vec{B}_0$.
The normal modes for such a case include light characteristics ($\omega_{\rm r} = k c$) in the short wavelength regime which turn into $ \omega_{\rm r} \sim \omega_{\rm p} $ for small values of $k$, where $\omega_{\rm p} = \sqrt{\omega_{\rm i}^2+\omega_{\rm e}^2}$ is the total plasma frequency.

\begin{figure}
\includegraphics[width=13.5cm]{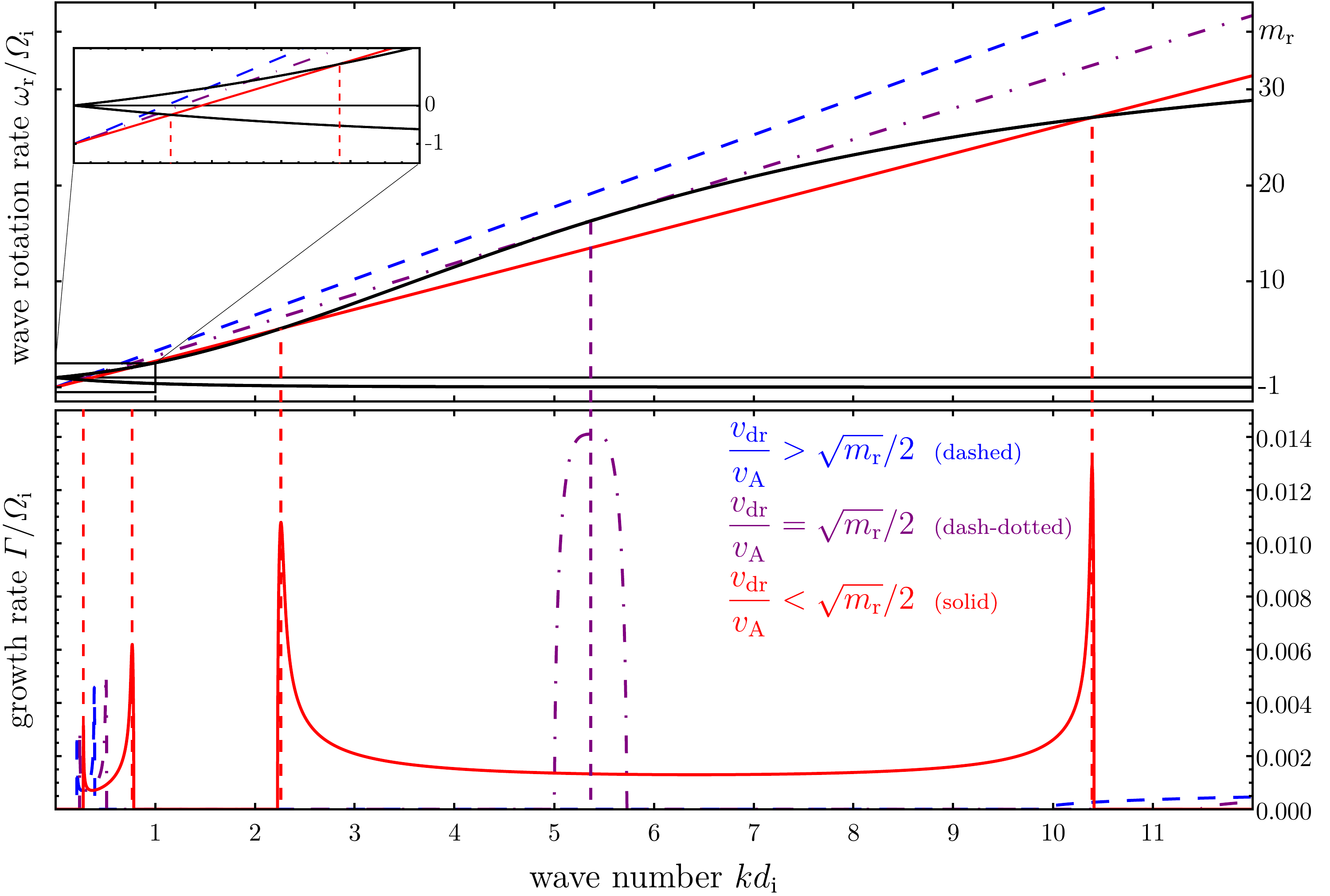}
\caption{%
Solutions of the dispersion relation with low-density drifting CRs in the rest frame of the background plasma  (Equation~\ref{eq:fulldisp}). We use the following parameters: $\varv_\A = 10^{-4} c$, $\alpha = n_{\rm cr}/n_0 = 10^{-6}$, $m_{\rm i}/m_{\rm e} = 36$, $v_{\perp, \rm e}=0$, and $\varv_{\perp,\rm i} = \varv_\A$.
The solutions are, in general, eight-complex values of $\omega$ at each wave mode $k$. In the top panel, we show the real values, $\omega_{\rm r}$, of three solutions: the CR ion cyclotron (coloured lines with different line styles, for three different parameter choices) as well as the background ion and electron cyclotron waves (black solid lines). We choose three values for the CR drift speed along $\vec{B}_0$ and indicate each case with a different colour and line style. In the bottom panel, we show the growth rate of the CR ion cyclotron solution, which is also the fastest growth rate, $\Gamma$, at each $k$, i.e., it has the largest imaginary part of $\omega$ of all solutions.
In the top panel, $\omega_{\rm r}$ represents the rate by which the electric and magnetic field perturbation vectors rotate. $\omega_{\rm r}>0$ ($\omega_{\rm r}<0$) indicates the same sense of rotation as the gyro motion of electrons (ions) around the large-scale background magnetic field $\vec{B}_0$.
Vertical dashed lines are solutions of Equation~\eqref{eq:kloc} and indicate values of the wave mode for which the CR ion-cyclotron wave in the background plasma rest frame is in exact resonance with the background cyclotron waves. These points exactly correspond to the locations of the peaks in the growth rate.
}
\label{fig:fig01}
\end{figure}

The other set of wave modes are the electron and ion cyclotron waves in the short wavelength regime, $k d_{\rm i} \ll 1$, where $d_{\rm i}=c/\omega_{\rm i}$ is the ion-skin depth.
More precisely, the electron-cyclotron branch includes the forward-propagating {\alf} waves ($\omega_{\rm r} = k \varv_\A$) for $k d_{\rm i} \ll 1 $, where $\varv_\A = B_0/\sqrt{ \mu_0  n_{\rm i} m_{\rm i}} = \Omega_{\rm i,0} c/\omega_{\rm i}$ is the {\alf} speed for ions.
This branch turns into whistler waves for shorter wavelengths at around $k d_{\rm i} > 1$, 
which become electron-cyclotron waves with $\omega_{\rm r} = |\Omega_{\rm e}|$ for $k d_{\rm e} = k d_{\rm i}/\sqrt{m_{\rm r}}  >1 $, where $d_{\rm e} = c /\omega_{\rm e}$ is the electron skin depth.
The ion-cyclotron branch is conceptually simpler: it includes the backward-propagating {\alf} waves for $k d_{\rm i} \ll 1$ and the ion-cyclotron wave modes with $\omega_{\rm r} = - \Omega_{\rm i}$ for $k d_{\rm i} \geq 1$.
An example for the electron and ion cyclotron branches is depicted by the black solid lines in the top panel of Figure~\ref{fig:fig01}.

\subsection{Resonances in the presence of drifting low density CRs}

Whether a drifting electron-ion plasma population (CR population) can excite resonant plasma instabilities is most easily seen by identifying the intersection points of background and CR wave modes in the $\omega$-$k$ plane: this yields solutions of wave modes for which the rotation rate of both background and CRs are in resonance.
Thus, it is for these wave modes that we expect the largest energy exchange between CRs and the background plasma populations.
In the following, we take a closer look at these resonant interactions in the case of CRs with a low number density, $n_{\rm cr}$, in comparison to the density of the background plasma $n_0$.
We characterise this by the CR-to-background number density ratio%
\footnote{%
The low values of $\alpha$ are motivated by the physical conditions in various astrophysical plasmas relevant for CR transport. For instance, GeV protons, which carry the majority of CR energy in our Galaxy, are characterised by $\alpha \sim 10^{-7}$ in the hot phase of the interstellar medium. A detailed discussion of how these estimates are obtained in various contexts is given in Appendix A and Figure 13 of \citet{sharp2}. The dependence of the growth rate of various CR-driven instabilities on $\alpha$ (in the limit of $\alpha \ll 1$) is also studied in detail by \citet{sharp2}, and a summary of the growth rates of these instabilities is provided in Table 1 of \citet{sharp2}.}%
$\alpha \equiv n_{\rm cr}/n_0 \ll 1 $ and adopt a mean relative drift speed $v_{\rm dr}$ of the CR population along the large scale background magnetic field.
In the background frame, one can derive the expected CR modes by solving the dispersion relation
\begin{eqnarray}
\label{eq:CRdisp}
\omega^2- k^2 c^2 + \zeta_{\rm e}^{-} (v_{\rm dr},0, \alpha n_0) + \zeta_{\rm i}^{-} (v_{\rm dr},0, \alpha n_0) = 0.
\end{eqnarray}
As anticipated, one solution of Equation~\eqref{eq:CRdisp} is the CR ion-cyclotron branch, which is comparable to that of the background, albeit with a Doppler-shifted rotation frequency, that is $\omega \rightarrow \omega + k \varv_{\rm dr}$.
Additionally, in the ion-cyclotron branch, the backward propagating {\alf}~waves starts to rotate at the Doppler-shifted ion-cyclotron frequency at $k d_{\rm cr,i} = k d_{\rm i}/ \sqrt{\alpha}\sim 1$.
In other words, within the ion-cyclotron branch, wave modes rotate with the Doppler-shifted ion-cyclotron frequency at $ k d_{\rm i}  \gtrsim \sqrt{\alpha}$, which is much less than unity. This means that, as observed from the background, this branch mainly produces waves rotating with Doppler-shifted ion-cyclotron frequency: $\omega_{\rm r} = - \Omega_{\rm i} + k \varv_{\rm dr}$.
We refer to this mode as the CR ion-cyclotron wave mode.

To verify that the individual wave modes of CRs and those of the background plasma are indeed the sum of the background and CR wave modes as seen in the background frame, we solve the full dispersion relation
\begin{eqnarray}
\label{eq:fulldisp}
\omega^2- k^2 c^2
+ \zeta_{\rm e}^{-} (0,0,  n_0) + \zeta_{\rm i}^{-} (0,0,n_0)
+ \zeta_{\rm e}^{-} (\varv_{\rm dr},\varv_{\perp, \rm e}, \alpha n_0) + \zeta_{\rm i}^{-} (\varv_{\rm dr},\varv_{\perp,\rm i}, \alpha n_0) = 0.
~~~~~
\end{eqnarray}
Note that in our setup, the zeroth-order current due to CR ions is compensated by CR electrons drifting at the same speed.
Here, we have added the possibility for CR populations to have a non-zero perpendicular velocity.
That is, CRs are distributed uniformly on a ring in the perpendicular velocity space, characterised by a radius of $\varv_{\perp}$.
Solutions for the full dispersion relation are shown in Figure~\ref{fig:fig01}, where we use a reduced mass ratio $m_{\rm r} = m_{\rm i}/m_{\rm e} = 36$ for visual purposes to reduce the separation between electron and ion cyclotron frequencies; $\sqrt{m_{\rm r}}/2 = 3$. We also adopt an ion {\alf} speed $\varv_\A$ at $10^{-4}c$, the density ratio $\alpha = n_{\rm cr}/n_0 = 10^{-6}$, $\varv_{\perp, \rm e}=0$, and $\varv_{\perp,\rm i}  = \varv_\A$. We use different values of $\varv_{\rm dr}/\varv_\A \approx \{3.7, 3, 2.7\}$, such that various interesting types of resonances are expected in the solutions of the dispersion relation. We note that this choice of parameters enables us to show the physics of resonance on a linear scale in the background plasma frame. In Section~\ref{sec:CR_frame}, we adopt more physically motivated parameters and show the resulting solutions of the dispersion relation in the CR drift frame.

For all values of $\varv_{\rm dr}$, we obtain the same exact background wave modes as described in the previous section, which are shown as black lines in the top panel of Figure~\ref{fig:fig01}. Differing values of $\varv_{\rm dr}$ result in various CR ion-cyclotron wave modes, which are shown as different colours and which exactly agree with $\omega_{\rm r} = - \Omega_{\rm i} + k \varv_{\rm dr}$ expected for the various cases.
In the bottom panel of Figure~\ref{fig:fig01}, we show the fastest growth rate obtained by solving the full dispersion relation in Equation~\eqref{eq:fulldisp}.
The real part of the most unstable modes are the CR ion-cyclotron waves that we show with different colours and line styles in the top panel of Figure~\ref{fig:fig01}. That is, in the unstable regions, only the rotating and propagating electromagnetic waves that are supported by the gyrating and drifting CRs are exponentially growing.
We note that in regions where the fastest growth rate is zero, any branch can be chosen, and we select the CR ion-cyclotron wave mode in this case.

To find the locations where the CR ion-cyclotron wave modes are in resonance with the background wave modes, we solve
\begin{eqnarray}
\label{eq:kloc}
[\omega^2- k^2 c^2 + \zeta_{\rm e}^{-} (0,0,n_0) + \zeta_{\rm i}^{-} (0,0,n_0)]_{\omega = -  \Omega_{\rm i} + k \varv_{\rm dr}} = 0, 
\end{eqnarray}
for wave mode $k$.
That is, the solutions are those wave modes for which the CR ion-cyclotron branch intersects the background wave modes in the cold HD limit.
This gives a fourth order polynomial which, in general, has four solutions for $k$ for any value $\varv_{\rm dr}/\varv_\A$, albeit these solutions may be degenerate.

For the case of $\varv_{\rm dr}/\varv_\A = 2.7 < \sqrt{m_{\rm r}}/2$ (solid-red curves in Figure~\ref{fig:fig01}), there exist four locations where the CR ion-cyclotron wave modes are in resonance with the background wave modes, i.e., Equation~\eqref{eq:kloc} has four distinct roots.
These expected locations for resonances are indicated with red-dashed vertical lines in the top and bottom panels of Figure~\ref{fig:fig01}: there are two resonances for $k d_{\rm i} <1$ and two resonances for $k d_{\rm i} >1$.
The two solutions for $k d_{\rm i} <1$ represent the intersections of the CR ion-cyclotron wave mode with both forward and backward propagating {\alf} waves, and thus resonance occurs approximately at $k d_{\rm i} = \varv_\A/(\varv_{\rm dr} \pm \varv_\A)$. This corresponds to the two resonant peaks in the growth rate due to the gyro-resonant (streaming) instability \citep{kulsrud+1969}.
The other two resonances at smaller scales at $k d_{\rm i} >1$ correspond to the two peaks in the growth rate due to the intermediate-scale instability \citep{sharp2}. In the next section, we show the growth rates when using realistic values of the mass ratio, for which the fastest growth rate due to the intermediate-scale instability is more than an order of magnitude greater then that at the gyroscales.

In Figure~\ref{fig:fig01}, the purple dash-dotted curves represent the case of $\varv_{\rm dr}/\varv_\A \sim  \sqrt{m_{\rm r}}/2$. We see that the two resonant peaks of the the intermediate-scale instability at $k d_{\rm i} >1$ merge into a single resonance. As in the previous case, the two peaks of the gyroscale instability for $k d_{\rm i} <1$ remain, but move closer together.
As $\varv_{\rm dr}/\varv_\A $ grows larger than $\sqrt{m_{\rm r}}/2$, the resonances at $k d_{\rm i} >1$ disappear, and thus, the intermediate-scale instability is no longer able to drive wave modes with $k d_{\rm i} > 1$ unstable (blue dashed curves in Figure~\ref{fig:fig01}).
That is, $\varv_{\rm dr}/\varv_\A \leq \sqrt{m_{\rm r}}/2$ is a condition for the possibility of resonance between CR ion-cyclotron wave modes and background wave modes at $k d_{\rm i} >1$, which is also the condition for the intermediate-scale instability.\footnote{The condition for the intermediate-scale instability $\varv_{\rm dr}/\varv_\A \leq \sqrt{m_{\rm r}}/2$ attains a small correction, i.e., growth may occur for slightly larger values of $\varv_{\rm dr}/\varv_\A$. However, this correction approaches zero for realistic values of $\varv_\A \ll 1$ and $m_{\rm r} \gg 1$. An analytical derivation of this correction can be obtained via Equation~\eqref{eq:kloc}.
}
The numerical identification of this instability condition was previously conducted by \citet{sharp2}.

It is important to acknowledge that in a realistic environment, background plasmas possess a finite temperature. Consequently, when incorporating this into the dispersion relation, it can influence wave modes occurring on scales smaller than the ion Debye length, which is much smaller in comparison to the ion skin depth for non-relativistic plasmas \citep{Reville+2008,Zweibel+Everett_2010}. Thus, it does not have any impact on the instabilities discussed in this paper. Furthermore, discussions and analysis of simulations in \citet{sharp2} and \citet{Lemmerz+2023} reveal that ion-cyclotron thermal damping has negligible effects on the driven wave modes.

\subsection{Instabilities in the rest frame of the CRs}
\label{sec:CR_frame}

In this section, we demonstrate that the intermediate-scale instability indeed drives CR comoving ion-cyclotron wave modes as found in \citet{sharp2}. To this end, we present the solution of the dispersion relation in the CR rest frame, where the background plasma is drifting with velocity $-\varv_{\rm dr}$ and hence drifting anti-parallel with respect to the background magnetic field $\vec{B}_0$.
The dispersion relation that we solve is given by:

\begin{eqnarray}
\label{eq:fulldisp-CR}
\omega^2- k^2 c^2
+ \zeta_{\rm e}^{-} (-\varv_{\rm dr},0,  n_0) + \zeta_{\rm i}^{-} (-\varv_{\rm dr},0,n_0)
+ \zeta_{\rm e}^{-} (0,\varv_{\perp, \rm e}, \alpha n_0) + \zeta_{\rm i}^{-} (0,\varv_{\perp, \rm i}, \alpha n_0) = 0.
\nonumber \\
\end{eqnarray}

\begin{figure}
\includegraphics[width=13.5cm]{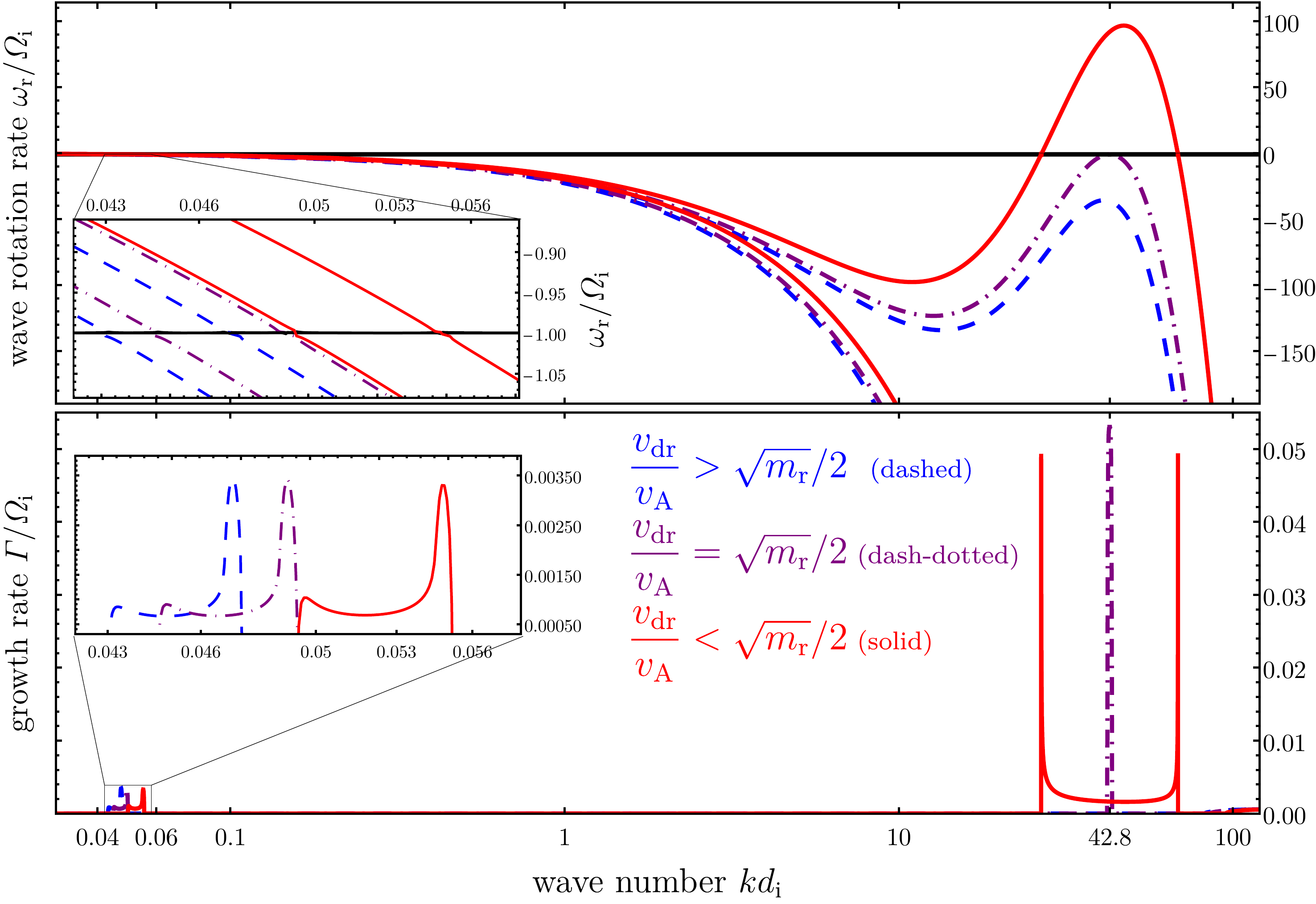}
\caption{\label{fig:fig02}%
Solutions of the dispersion relation of low-density CRs in the rest frame of the CRs, i.e., using Equation~\eqref{eq:fulldisp-CR}, with the following parameters: $\varv_\A = 10^{-4} c$, $\alpha = n_{\rm cr}/n_0 = 10^{-6}$, $m_{\rm i}/m_{\rm e} = 1836$, $v_{\perp,\rm e}=0$, and $\varv_{\perp, \rm i} = \varv_\A$.
We vary the relative drift speed (the background plasma is drifting with $\varv_{\rm dr}$ anti-parallel to background magnetic field $\vec{B}_0$) to obtain similar cases as in Figure~\ref{fig:fig01} but with a realistic ion-to-electron mass ratio.
In the top panel, we show the rotation rate of wave modes, with the solid-black line representing the CR ion-cyclotron wave mode which is the same for all cases.
The background wave modes are the same in the background rest frame, however as seen in the rest frame of the CRs, the rotation and growth rates of the background waves are different and are thus indicated with various colours and line styles.
In the bottom panel, we show the growth rates of the CR ion-cyclotron waves in all cases, and we find that these are, also, the fastest-growing rates for the instability.
This shows that, in the case of realistic $m_{\rm r}$, the dominance of intermediate-scale growth rate compared to that at the gyroscale ($kd_{\rm i}<1$) is much more pronounced and that the growth rate for the forward propagating {\alf}~waves is also larger in comparison to the backward propagating wave.
}
\end{figure}

Figure~\ref{fig:fig02} shows some solutions of Equation~\eqref{eq:fulldisp-CR} for $\varv_\A = 10^{-4} c$, $\alpha = n_{\rm cr}/n_0 = 10^{-6}$,\footnote{The low values of $\alpha$ are motivated by the physical conditions in various astrophysical plasmas relevant for CR transport as discussed in Appendix~A of \citet{sharp2}.} $m_{\rm r} = m_{\rm i}/m_{\rm e} = 1836$, $v_{\perp, \rm e}=0$, and $\varv_{\perp, \rm i} = \varv_\A$, along with various background plasma drift speeds such that $\varv_{\rm dr}/\varv_\A \sim \{22.28, 21.45 , 19.281 \} \sim \{1.05, 1, 0.9 \} \sqrt{m_{\rm r}}/2$.
In the top panel of that figure, the CR ion-cyclotron wave modes are shown as a black line for different values of $\varv_{\rm dr}$. Meanwhile, the background cyclotron wave modes are rotated differently depending on the drift speed, leading to intersections at both long ($kd_{\rm i} <1$) and short ($kd_{\rm i} >1$) wavelengths for $\varv_{\rm dr}/\varv_\A \leq \sqrt{m_{\rm r}}/2$. These intersections lead to instabilities at these wave modes.
The bottom panel of Figure~\ref{fig:fig02} shows the growth rates of the ion-cyclotron wave modes, which are also the fastest growing modes. This demonstrates that the driven modes are ion-cyclotron waves at all wavelengths in the rest frame of CRs.
If the condition for the intermediate-scale instability is not fulfilled ($\varv_{\rm dr}/\varv_\A > \sqrt{m_{\rm r}}/2$), the instability no longer operates and short wavelengths modes are stable.

The solutions in Figure~\ref{fig:fig02} show that, for realistic values of $m_{\rm r}$, the intermediate-scale instability growth rates significantly dominate over gyroscale growth rates.
This dominance is even more pronounced at larger pitch angles because the growth rate at the peaks of the intermediate-scale instability is proportional to $(\varv_{\perp,\rm i}/\varv_\A)^{2/3}$ \citep{sharp2}.
Additionally, it is important to note that the growth rate is higher for the resonance with forward-propagating {\alf}~waves at the gyroscale ($kd_{\rm i}<1$) compared to that with the backward-propagating waves.

\section{Instabilities with approximate background plasma descriptions}
\label{sec:Bgs}

In this section, we revisit the dispersion relation in the background plasma rest-frame to investigate the effect of approximating the background plasma description on the nature of the emerging instabilities.
One commonly used approximation for the background plasma dispersion assumes the MHD dispersion relation, where it is assumed that $\omega \ll \Omega_{\rm i} \ll |\Omega_{\rm e}|$. In this limit, the third and fourth terms in Equation~\eqref{eq:fulldisp} are reduced to $\omega^2 c^2/\varv_\A^2$ and the first term ($\omega^2$) is neglected, due to the fact that $
\varv_\A^2 \ll c^2 \Rightarrow 
\omega^2 \ll \omega^2 c^2/\varv_\A^2$ . The dispersion relation is for
\begin{eqnarray}
{\rm MHD:~~~~} &&
\frac{\omega^2 c^2}{\varv_\A^2}- k^2 c^2
+ \zeta_{\rm e}^{-} (\varv_{\rm dr},\varv_{\perp,\rm e}, \alpha n_0) + \zeta_{\rm i}^{-} (\varv_{\rm dr},\varv_{\perp, \rm i}, \alpha n_0) = 0.
~~~~~
\label{eq:fulldisp-mhd}
\end{eqnarray}
This approximation is typical in works aiming at computing various types of CR driven instabilities \citep[see, e.g.,][]{Bell2004,Zweibel-2003,Amato+2009,Bai2019}.
Another possible approximation involves considering the impact of finite but small values of $k d_{\rm i}$, which are wave modes comparable to the ion skin-depth, as is typically done in the Hall-MHD approximation. In this case, the dispersion relation approximates the behaviour of {\alf} and whistler waves found in the cold HD dispersion relation. Within the Hall-MHD approximation, the background plasma contributions \citep[see, e.g., Section 14.4.4 of][]{goedbloed_keppens_poedts_2010} are reduced to $\omega^2 c^2/[\varv_\A^2 (1+ k d_{\rm i})^2]$, leading to the following modified dispersion relation in
\begin{eqnarray}
{\rm Hall{\text -}MHD:~~~~} &&
\frac{\omega^2 c^2}{\varv_\A^2 (1+ k d_{\rm i})^2 }- k^2 c^2 
+ \zeta_{\rm e}^{-} (\varv_{\rm dr},\varv_{\perp, \rm e}, \alpha n_0) + \zeta_{\rm i}^{-} (\varv_{\rm dr},\varv_{\perp,\rm i}, \alpha n_0) = 0.
~~~~~
\label{eq:fulldisp-hall}
\end{eqnarray}

\begin{figure}
\includegraphics[width=13.5cm]{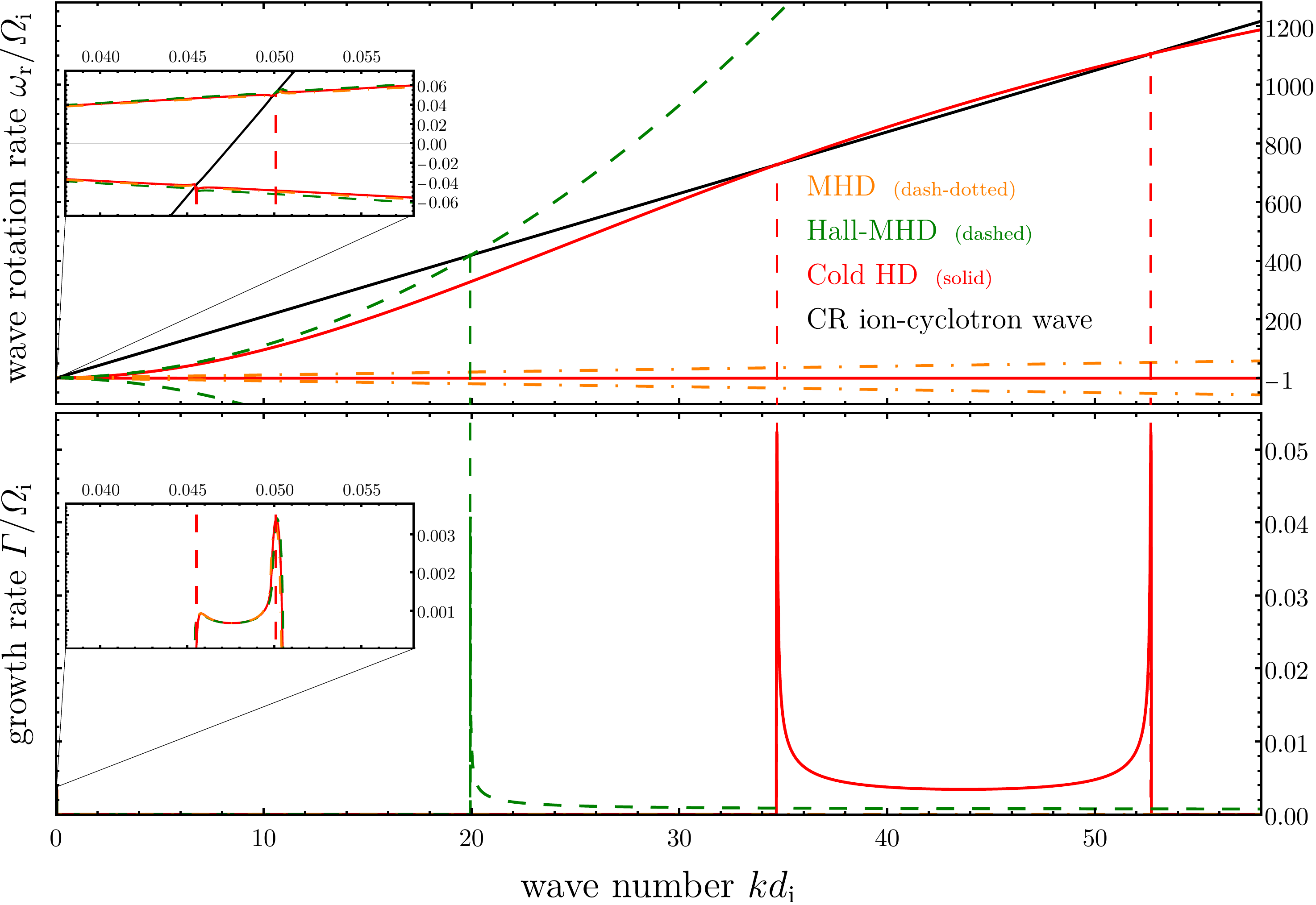}
\caption{\label{fig:fig03}%
Solutions of the dispersion relation for various approximations of the background plasma using the following parameters: $\varv_\A = 10^{-4} c$, $\alpha = n_{\rm cr}/n_0 = 10^{-6}$, $m_{\rm i}/m_{\rm e }= 1836$, $v_{\perp, \rm e}=0$, $\varv_{\perp,\rm i} = \varv_\A$ and $\varv_{\rm dr}/\varv_\A = 0.98 \sqrt{m_{\rm r}}/2 \sim 20.99$.
The top panel shows the rotation rate of various wave modes, while the bottom panel shows the growth rate of the corresponding unstable CR ion-cyclotron wave mode, which has the fastest growth rate.
The black line in the top panel represents the CR ion-cyclotron wave mode, which follows the same dispersion in all cases, namely $\omega_{\rm r} = k \varv_{\rm dr} - \Omega_{\rm i}$.
In the MHD case, the background wave modes follow a dispersion relation of $\omega = \pm k \varv_\A$ (orange curves in the top panel), while in the Hall-MHD case, the background waves follow a dispersion relation of $\omega = \pm k \varv_\A (1+k d_{\rm i})$.
The bottom panel reveals that the growth rates of the CR ion-cyclotron wave mode at the gyroscale ($k d_{\rm i}<1$) are almost identical for different approximations of the background plasma.
However, at intermediate scales where $k d_{\rm i}>1$, the MHD background approximation fails to capture the fastest growing instability, namely the intermediate-scale instability with growth rates shown by the red curve.
In contrast, the Hall-MHD approximation reproduces the first peak of the  dominant instability growth rate, although at a reduced rate and at a larger wavelength.
For all wavelengths shorter than this peak, the use Hall-MHD approximation leads to wrong growth rates.
Vertical lines indicate the predicted intersection points of CR ion-cyclotron waves and background wave modes for the different approximations.
}
\end{figure}

In Figure~\ref{fig:fig03}, we show the solutions of the dispersion relations with various background plasma assumptions, i.e., the solutions for Equations~(\ref{eq:fulldisp}, \ref{eq:fulldisp-mhd}, and \ref{eq:fulldisp-hall}) using the following parameters: $\varv_\A = 10^{-4} c$, $\alpha = n_{\rm cr}/n_0 = 10^{-6}$, $m_{\rm i}/m_{\rm e} = 1836$, $v_{\perp, \rm e}=0$, $\varv_{\perp, \rm i} = \varv_\A$, and $\varv_{\rm dr}/\varv_\A = 0.98 \sqrt{m_{\rm r}}/2 \approx 20.99$.
In all cases, we fix the CR drift speed, resulting in the same CR ion-cyclotron wave mode (shown as a black line in the top panel) that follows the expected dispersion relation $\omega_{\rm r} = k \varv_\A - \Omega_{\rm i}$.
When using the MHD background approximation, the background wave modes follow the expected dispersion relation $\omega_{\rm r} = \pm k \varv_\A$ (shown as orange dash-dotted curves in the top panel).
In the case of Hall-MHD, the dispersion of background wave modes, as expected, follows $\omega_{\rm r} = \pm k \varv_\A (1+k d_{\rm i})$; shown as green dashed curves in the top panel.

In the bottom panel of Figure~\ref{fig:fig03}, we show the growth rates of the CR ion-cyclotron waves (which are the fastest growth rates) for different approximations of the background plasma.
At large (gyro) scales, all approximations produce similar growth rates peaking at the same wave modes, indicating that using any of these approximations for the background plasma leads to correctly capturing the gyroscale instability.
However, at smaller scales, i.e., scales where the intermediate-scale instability operates ($kd_{\rm i}>1$), the MHD approximation of the background plasma dispersion wrongly predicts complete stability at these scales. In the case of Hall-MHD, the growth of the first peak of the intermediate-scale instability is approximately reproduced but at a longer wavelength.
For shorter wavelengths, the use of the Hall-MHD approximation predicts completely wrong growth rates compared to those found when using the cold-HD dispersion for the background plasma.
An additional  worrisome implication arising from the use of the Hall-MHD approximation is its incorrect prediction that shorter wavelength modes are universally unstable, even in cases where the condition for driving the intermediate-scale instability is not satisfied.

\section{Nature of CR scattering at different resonant scales}
\label{sec:scattering}

So far, we have focused on the influence of gyrotropic CR populations in driving perpendicular electromagnetic perturbations. In the absence of such perturbations, particles follow trajectories characterised by a constant drift speed along the direction of the magnetic field vector, while simultaneously gyrating at their gyrofrequency.
In this section, our goal is to evaluate the effects of these perturbations on the particle trajectories, specifically how particles are scattered by the induced parallel electromagnetic perturbations. To accomplish this, we examine the Lorentz force acting on a particle with velocity $\bvarv =\{ \varv_{x} , \varv_{y}, \varv_{z} \}$, charge $q_\rmn{s}$, and mass $m_\rmn{s}$ caused by these perturbations.
In Fourier space, along the particle trajectory, the momentum equation is:
\begin{eqnarray}
\label{eq:lfk}
\frac{\rmn{d} \gamma \bvarv}{\rmn{d}t} =
\frac{q_\rmn{s}}{m_\rmn{s}} 
\left( \delta \vec{E}^k + \bvarv \bs{\times} \delta \vec{B}^k \right)
=
\frac{q_\rmn{s}}{m_\rmn{s}} 
\begingroup
\renewcommand*{\arraystretch}{1.45}
\begin{bmatrix}
\varv_{y}  \delta B^{k}_{z} - \varv_{z} \delta  B^{k}_{y}
\\
-(\varv_{x}-\varv_{\rm ph}) \delta B^{k}_{z}
\\
~~ (\varv_{x}-\varv_{\rm ph})\delta  B^{k}_{y} 
\end{bmatrix}
\endgroup
\end{eqnarray}
Here, we use the fact that $\vec{k} \parallel \vec{B}_0$ and both vectors are aligned with $\bs{\hat{x}}$. Furthermore, we utilise the relation $\vec{k} \bs{\times} \delta \vec{E}^k = \omega \delta \vec{B}^k$ to find $\delta \vec{E}^k = - \bvarv_{\rm ph} \bs{\times} \delta \vec{B}^k$. This indicates that particle scattering in the parallel direction is significantly influenced by the phase difference between the particle's perpendicular velocities and the magnetic field perturbations. On the other hand, scattering of particles in the perpendicular direction crucially depends on the difference between the relative drift speed of the particles and the waves.

In previous sections, we found that the propagation of CRs destabilise waves at different wavelengths. The real frequency of the unstable wave modes, in the rest frame of the background, is always given by $\omega_{\rm r} = k \varv_{\rm dr} - \Omega_{\rm i}$. 
Consequently, the phase velocity of the unstable waves can be expressed as $\varv_{\rm ph} = \varv_{\rm dr} - \varv_\A/(k d_{\rm i})$.
At the gyroscale ($k d_{\rm i} < 1$), wave growth peaks at wave modes where $k d_{\rm i} \approx \varv_\A/(\varv_{\rm dr} \mp \varv_\A)$. This leads to a phase velocity of approximately $\varv_{\rm ph} \approx \pm \varv_\A$ for forward ($+$) and backward ($-$) propagating {\alf}~waves.
On the other hand, the fastest growth due to the intermediate-scale instability occurs for $k d_{\rm i} > 1$. Since typically $\varv_{\rm dr} \gg \varv_\A$, the phase velocity of the driven unstable modes is approximately $\varv_{\rm ph} \approx \varv_{\rm dr}$.
To summarise, the phase speed of the growing wave modes can be expressed as follows:
\begin{equation}
\label{eq:vph}
\varv_{\rm ph} =
\left\{
\begin{aligned}
&\pm \varv_\A, \quad &kd_{\rm i} \ll 1, \\
&\varv_{\rm dr}, \quad &kd_{\rm i} > 1.
\end{aligned}
\right.
\end{equation}

Therefore, although both the gyro and intermediate-scale instabilities are resonant instabilities, they lead to perturbations with distinct phase speeds. Due to the significant disparity in phase velocity between these two scales, the scattering of CR ions
by these electromagnetic perturbations exhibits notable differences, as can be seen from Equation~\eqref{eq:lfk}.
The wave modes driven by the intermediate-scale instability scatter particles that drive them only in the direction parallel to $\vec{B}_0$, since $\varv_{x} = \varv_{\rm dr} \sim \varv_{\rm ph}$.
However, particles with different parallel drift speeds can be scattered both in parallel and perpendicular directions due to wave modes driven by the intermediate-scale instability.
On the other hand, scattering occurs in both parallel and perpendicular directions at the gyroscale. It can be demonstrated from Equation~\eqref{eq:lfk} that the gyroscale waves result in energy-conserving scattering in the frame of the driven {\alf} waves \citep{sharp2}.

\section{Summary}
\label{sec:summary}

This paper examines the physics of resonant instabilities driven by CR ions with a gyrotropic momentum distribution. These instabilities occur most rapidly when a resonance between the Doppler-shifted background and CR wave modes occurs. The Doppler shift arises from the relative drift between the background and CR plasma.
The relative drift sets the location of resonances and hence the most unstable wavelengths of various instabilities. It leads to two important resonant instabilities: the gyroscale instability at large scales \citep{kulsrud+1969} and the recently found intermediate-scale instability \citep{sharp2}.

The gyroscale instability has peak growth rates when the CR ion-cyclotron wave mode resonates with both, forward and backward propagating {\alf} waves of the background plasma.
In the background frame, the resonances occur when $- \Omega_{\rm i} + k \varv_{\rm dr} \rightarrow \pm k \varv_\A$, which results in the most unstable wave modes at $k_{\rm g,i}^{\pm}d_{\rm i} = \varv_\A /(\varv_{\rm dr} \mp \varv_\A)$.
From Figure~\ref{fig:fig02}, it is evident that resonance with the forward-propagating {\alf} wave leads to faster growth at larger wave modes $k_{\rm g,i}^{+}$, thus it is called the ion-gyro scale. 
Since $\varv_{\rm dr} \gg \varv_\A$, both growth peaks at the gyroscales occur for wavelengths larger than the ion-skin depth, i.e., $k_{\rm g,i}^{\pm}d_{\rm i} <1$.
Therefore, in the linear regime, the gyroscale instability can be accurately described when approximating the electron-ion background plasma using either MHD or Hall-MHD approximations (see Figure~\ref{fig:fig03}).

For an electron-ion background plasma, additional resonances occur at shorter wavelengths, where $k d_{\rm i} > 1$. These resonances give rise to the intermediate-scale instability that destabilise wave modes between the ion gyroscale, $k_{\rm g,i}^{+}$ and the electron gyroscale, $k_{\rm g,e} = m_{\rm r} k_{\rm g,i}^{+}$, where $m_{\rm r}$ represents the ion-to-electron mass ratio.
At these resonant scales, the peak growth rates of the intermediate-scale modes are significantly larger compared to those at the gyroscales.
This establishes the intermediate-scale instability as the fastest instability in the linear regime of the resonant CR-driven instabilities.
Moreover, the dominance of the peak growth of the intermediate-scale instability is further amplified when CR ions possess larger pitch angles, corresponding to a larger perpendicular velocity ($\varv_{\perp}$) \citep[see Table 1 of][]{sharp2}.

As depicted in Figure~\ref{fig:fig03}, the utilisation of the MHD approximation for describing the linear response of the background plasma hinders the occurrence of resonances at short wavelengths. Consequently, such an approximation suppresses the dominant, intermediate-scale instability.
Conversely, employing the Hall-MHD approximation for the background plasma captures a resonance at $k d_{\rm i} > 1$, and, while the fastest growth rate associated with the intermediate-scale instability is approximately captured, it is associated with a wrong wave number. Additionally, this approximation erroneously predicts the presence of intermediate-scale instability even when the conditions for the instability are not met. 

We argue in Section~\ref{sec:intro} that background temperatures are unlikely to impact the growth of the instabilities addressed in this paper. Supporting evidence for this assertion can be found in the simulations conducted by \citet{sharp2}, where the background plasma was characterised by high temperatures, yet exhibited excellent agreement with the growth rates predicted from the dispersion relation assuming a cold background plasma.
On the other hand, it is natural to contemplate whether these instabilities, particularly the newly discovered intermediate-scale instability, persist under different and potentially more realistic velocity distributions for CR ions. The fundamental explanation for the origin of these instabilities lies in the resonance between the ion-cyclotron wave modes of CR ions and the background wave modes.
Consequently, any velocity distribution that supports CR ion cyclotron modes while adhering to the instability conditions will excite this instability.
This argument is supported by circumstantial evidence from particle-in-cell simulations of shocks conducted by \citet{Shalaby+2022ApJ}, where the instability is clearly driven by CRs with a thermal velocity distribution.

Thus, while our choice of the CR ion distribution ensures the ease of repeatability in our analytical calculations, it does not imply that the instability is exclusively associated with such a choice of CR ion velocity distribution.
The choice of  a particular CR ion velocity distribution could impact the growth rates of different instabilities. However, qualitative considerations show that it is highly likely that the intermediate-scale instabilities will remain dominant even in these cases.
An analytical demonstration of this is, however, deferred to future studies.

\section{Outlook}
\label{sec:outlook}

The presence of the intermediate-scale instability is vital for the efficiency of electron acceleration in parallel electron-ion non-relativistic shocks as seen in fully kinetic particle-in-cell simulations \citep{sharp,sharp2}.
Moreover, when the condition for the instability is not met in simulations, a notable decrease in the efficiency of electron acceleration is observed \citep{Shalaby+2022ApJ}.
That is, the significance of this instability extends beyond its potential to regulate the transport of CRs in various astrophysical scenarios.
It could substantially influence particle injection and acceleration processes occurring at shocks, the escape of CRs from their sources into the interstellar medium, and CR-driven galactic winds.
This novel understanding of the fundamental physics underlying CR-driven resonant plasma instabilities will contribute to elucidate the critical role played by CRs in many astrophysical environments.

\section*{Funding}
The authors acknowledge support by the European Research Council under ERC-AdG grant PICOGAL-101019746.

\section*{Author ORCID}

M.\ Shalaby, \href{https://orcid.org/0000-0001-9625-5929}{https://orcid.org/0000-0001-9625-5929}

T.\ Thomas, \href{https://orcid.org/0000-0002-7443-8377}{https://orcid.org/0000-0002-7443-8377}

C.\ Pfrommer, \href{https://orcid.org/0000-0002-7275-3998}{https://orcid.org/0000-0002-7275-3998}

R.\ Lemmerz, \href{https://orcid.org/0000-0002-4683-8517}{https://orcid.org/0000-0002-4683-8517}

V.\ Bresci, \href{https://orcid.org/0000-0001-7237-3373}{https://orcid.org/0000-0001-7237-3373}

\section*{Declaration of interests}
The authors report no conflict of interest.

\bibliography{refs}
\bibliographystyle{jpp}

\end{document}